\def\kms{\ifmmode{\,\hbox{km}\,s^{-1}}\else {\rm\,km\,s$^{-1}$}\fi}

\def\kmsm{{\rm\,km\,s^{-1}\,Mpc^{-1}}}

\def\hmpc{\ifmmode{h^{-1}\,\hbox{Mpc}}\else{$h^{-1}$\thinspace Mpc}\fi}
\def\hkpc{\ifmmode{\,h^{-1}\,{\rm kpc}}\else {$h^{-1}$\,kpc}\fi}

\def\et{{\it et~al.}~}

\documentclass{aastex}

\begin{document}

\title{The Merger Rate to Redshift One from Kinematic Pairs: \\
Caltech Faint Galaxy Redshift Survey XI}

\author{R.~G.~Carlberg\altaffilmark{1,2},
Judith~G.~Cohen\altaffilmark{3},
D.~R.~Patton\altaffilmark{1,2},
Roger~Blandford\altaffilmark{4},
David~W.~Hogg\altaffilmark{5,6},
H.~K.~C.~Yee\altaffilmark{1,2},
S.~L.~Morris\altaffilmark{1,7},
H.~Lin\altaffilmark{1,6,8},
Lennox L. Cowie\altaffilmark{9,10}, 
Esther Hu\altaffilmark{9,10},
and
Antoinette Songaila\altaffilmark{9,10}
}

\altaffiltext{1}{Visiting Astronomer, Canada--France--Hawaii Telescope, 
        which is operated by the National Research Council of Canada,
        le Centre National de Recherche Scientifique, and the University 
	of Hawaii.}
\altaffiltext{2}{Department of Astronomy, University of Toronto, 
        Toronto ON, M5S~3H8 Canada}
\altaffiltext{3}{Department of Astronomy, Caltech 105-24, 
	Pasadena, CA 91125}
\altaffiltext{4}{Theoretical Astrophysics, Caltech 130-33,
	Pasadena, CA 91125}
\altaffiltext{5}{Institute for Advanced Study, Olden Lane,
	Princeton, NJ 08540}
\altaffiltext{6}{Hubble Fellow}
\altaffiltext{7}{Dominion Astrophysical Observatory, 
        Herzberg Institute of Astrophysics,    ,  
        National Research Council of Canada,
        5071 West Saanich Road,
        Victoria, BC, V8X~4M6, Canada}
\altaffiltext{8}{Steward Observatory, University of Arizona,
        Tucson, AZ, 85721}
\altaffiltext{9}{Visiting Astronomer, W. M. Keck Observatory, jointly
	operated by the California Institute of Technology and the
	University of California.}
\altaffiltext{10}{Institute for Astronomy, University of Hawaii,
	2680 Woodlawn Drive, Honolulu, HI 97822}

\begin{abstract} 
The rate of mass accumulation due to galaxy merging depends on the
mass, density, and velocity distribution of galaxies in the near
neighborhood of a host galaxy. The fractional luminosity in kinematic
pairs combines all of these effects in a single estimator which is
relatively insensitive to population evolution. Here we use a
k-corrected and evolution compensated volume-limited sample having an
R-band absolute magnitude of $M_R^{k,e} \le -19.8+5\log{h}$ mag
drawing about 300 redshifts from CFGRS and 3000 from CNOC2 to measure
the rate and redshift evolution of merging.  The combined sample has
an approximately constant co-moving number and luminosity density from
redshift 0.1 to 1.1 ($\Omega_M=0.2, \Omega_\Lambda=0.8$); hence, any
merger evolution will be dominated by correlation and velocity
evolution, not density evolution. We identify kinematic pairs with
projected separations less than either 50 or 100 \hkpc\ and rest-frame
velocity differences of less than 1000\kms.  The fractional luminosity
in pairs is modeled as $f_L(\Delta v,r_p,M_r^{ke})(1+z)^{m_L}$ where
$[f_L,m_L]$ are $[0.14\pm0.07,0\pm1.4]$ and $[0.37\pm0.7,0.1\pm0.5]$
for $r_p\le 50$ and 100\hkpc, respectively ($\Omega_M=0.2,
\Omega_\Lambda=0.8$). The value of $m_L$ is about 0.6 larger if
$\Lambda=0$.  To convert these redshift space statistics to a merger
rate we use the data to derive a conversion factor to physical space
pair density, a merger probability and a mean in-spiral time. The
resulting mass accretion rate per galaxy ($M_1,M_2\ge 0.2 M_\ast$) is
$0.02\pm0.01(1+z)^{0.1\pm0.5} M_\ast~{\rm Gyr}^{-1}$.  Present day
high-luminosity galaxies therefore have accreted approximately
$0.15M_\ast$ of their mass over the approximately 7 Gyr to redshift
one.  Since merging is likely only weakly dependent on host mass, the
fractional effect, $\delta M/M \simeq 0.15M_\ast/M$, is dramatic for
lower mass galaxies but is, on the average, effectively perturbative
for galaxies above $M_\ast$.
\end{abstract}


\section{Introduction}

Merging is a fundamental mode of stellar mass addition to galaxies.
Moreover, merging brings in new gas and creates gravitational
disturbances that enhance star formation or fuel a nuclear black hole.
The general process of substructure infall may be the rate fixing
process for the buildup of a galaxy's stars and consequently may
largely regulate its luminosity history.  Gravitational forces on
relatively large scales dominate merger dynamics which allows direct
observation of the mechanism, although with the considerable
complication that dark matter dominates the mass.  N-body simulations
\citep{tt,bh} give the detailed orbital evolution, morphological
disturbances and eventual outcomes of the encounters of pairs of
galaxies.

The purpose of this paper is to estimate the rate of mass gain per
galaxy due to mergers over the redshift zero to one interval.  Our
primary statistic is the fractional luminosity in close kinematic
pairs, which is readily related to n-body simulations and sidesteps
morphological interpretation. This approach provides a clear sample
definition which is closely connected to the large scale dynamics of
merging. In common with all merger estimates it requires an estimate
of the fraction of the pairs that will merge and a mean time to
merger.

The number of kinematic pairs is proportional to the volume integral
at small scales of the product of two-point correlation function,
$\xi$, and the luminosity function (LF).  The high luminosity
galaxies appear to be evolving purely in luminosity
\citep{cfrs_lf,huan_lf}, which can be easily compensated. 
The measured evolution of $\xi$ suggests that the density of physical
pairs should not vary much with redshift, $(1+z)^{0\pm1}$
\citep{cfrsxi,kkeck,cnoc_xi}. This inference is in notable contrast with the
pair counts or morphological typing approaches to merger estimation
\citep{zk,cpi,ye_pairs,patton_cnoc1,cfrs_mg}, which suggest that
merging rate by number varies as $(1+z)^{3\pm1}$. HST
photometric pairs, with no redshift information leads to 
a dependence of  $(1+z)^{1.2\pm0.4}$ \citep{mdss}.

In the next section we combine the Caltech Faint Galaxy Redshift
Survey (CFGRS) and the Canadian Network for Observational Cosmology
field galaxy survey (CNOC2) from which we construct evolution
compensated, volume-limited, subsamples. In Section 3 we measure the
fractional luminosity in 50 and 100\hkpc\ companions as a function of
redshift. The CNOC2 sample is used in Section 4 to relate this wide
pair sample to a close pair sample which is more securely converted
into a mass merger rate. Section 5 discusses our conclusions. We
use $H_0= 100h\kmsm$, $\Omega_M=0.2$ in open and flat cosmologies.

\section{The CFGRS and CNOC2 Volume-Limited Samples}

The CFGRS sample of the HDF plus flanking fields is discussed in
detail elsewhere \citep{cfgrs_phot,cfgrs_z}. We use the high coverage
subsample lying within a 240 arcsecond radius circle, with a center
located at 12$^h$ 36$^m$ 50$^s$ and 62$^\circ$ 12$^\prime$
55$^{\prime\prime}$ (J2000). The computed magnitude selection
function, $s(m_R)$, (in Cousins R) is accurately approximated as a
constant 90\% spectroscopic completeness for $m_R<22.8$ mag with a
linear decline to 19\% at $m_R<23.4$ mag, our sample limit.  The
magnitude weight is $1/s(m_R)$.  The CFGRS k-corrections and evolution
compensation are here approximated as $k(z)= K z$ mag from the tables
of Poggianti (1997). For galaxies that Cohen et al. (2000) classify as
``E'' (emission), $K=1.0$, ``A'' types have $K=2.0$ and all types have
$K=1.7$

The CNOC2 selection weights and k-corrections are discussed in Yee et
al. (2000).  .  The evolution of the luminosity function is
approximated as a uniform $M_\ast(z) = M_\ast-Qz$, with $Q\simeq 1$,
\citep{huan_lf} which we use over the entire CNOC2-CFGRS redshift range.

The kinematic pair fraction is directly proportional to the mean
density of the sample and is therefore sensitively dependent on
correcting to a complete and uniform sample \citep{patton_ssrs2}.  The
most straightforward approach is to impose a strict volume limit.  For
our primary sample we will limit the CFGRS and the CNOC2 samples at
$M_R^{k,e} = -19.8+5\log{h}$ mag, which yields volume-limited samples
of about 300 CFGRS galaxies between redshift 0.3 to 1.1 and 3000 CNOC2
galaxies between 0.1 to 0.5. The volume density of the sample is
approximately constant at $1.2\times10^{-2} h^3$ Mpc$^{-3}$ over the
entire redshift range for $\Omega_\Lambda=0.8$ but rises roughly as
$(1+z)^{0.8}$ for $\Omega_\Lambda=0$. Both the CFGRS and the CNOC2
surveys are multiply masked, which minimizes the effects of slit
crowding, however there is still a measurable pair selection function.
The CNOC2 catalogue has about a 20\% deficiency of close angular
pairs. We model the measured angular pair selection weight as,
$w(\theta) =[1+a_s {\rm tanh}(\theta/\theta_s)]^{-1},$ where
$[a_s,\theta_s]$ is $[0.5,5^{\prime\prime}]$ for the CFGRS sample and
$[-0.3,10^{\prime\prime}]$ for the CNOC2 sample with typical pair
corrections being 10\%.

\section{The Pair Fractional Luminosity Fraction}

The preferred choice of pair statistic depends on the application
\citep{patton_ssrs2}. Here we are primarily interested in the impact
of merging on galaxy mass increase, for which the k-corrected,
evolution compensated R luminosity is a stand-in. The rate of merging
per galaxy depends on the density of galaxies in the near
neighborhood and their velocity distribution. As a practical redshift
space estimator, we compute the fractional luminosity in close kinematic pairs,
\begin{equation}
f_L(z|\Delta v^{\rm max},r_p^{\rm max}, M_R^{k,e}) = {{\sum_j 
	\sum_{i\ne j,<\Delta v^{\rm max},<r_P^{\rm max}} 
	w_j  w_i w(\theta_{ij})
	L_i }\over
	{\sum_j w_j L_j}},
\label{eq:fl}
\end{equation}
where the weights, $w_i$, allow for the magnitude selection function.
Note that the $ij$ and $ji$ pairs are both counted.  The ratio has the
benefit of being fairly stable for different luminosity limits,
self-normalizing for luminosity evolution, identical to a mass ratio
for a fixed $M/L$ population. For an unperturbed pair luminosity
function it is mathematically identical to the $N_c$ of Patton \et\
(2000a) although constructed out of somewhat different quantities. The
two parameters, $\Delta v^{\rm max}$ and $r_p^{\rm max}$, are chosen
on the basis of merger dynamics and the characteristics of the
sample. The rate of mass increase per galaxy is calculated from this
statistical estimator using a knowledge of merger dynamics and the
measured correlations and kinematics of galaxy pairs in the sample.

The mean fractional pair luminosity, based on 18 CFGRS pairs and 91
CNOC2 pairs, with $\Delta v\le 1000
\kms$ and $5\le r_p \le 50\hkpc$ pairs is displayed in Figure~1. 
These kinematic separation parameters are larger than is suitable for
reliably identifying ``soon-to-merge'' pairs. However, they provide a
statistically robust connection to those pairs and take into account
the lower velocity precision and sample size of the CFGRS relative to
CNOC2.  The errors are computed from the pair counts,
$n_p^{-1/2}$. The measurements of $f_L(z)$ in Figure~1 are fit to
$f_L(\Delta v,r_p,M_r^{ke})(1+z)^{m_L}$, finding $[f_L,m_L]$ of
$[0.14\pm0.07,0\pm1.4]$ for $r_p \le 50\hkpc$ pairs and
$[0.37\pm0.7,0.1\pm0.5]$ for $ r_p\le 100\hkpc$, both for
$\Omega_M=0.2, \Omega_\Lambda=0.8$. The increase with $r_p$ of $f_L$
is consistent with a $\gamma=1.8$ two-point correlation function. If
$\Omega_M=0.3, \Omega_\Lambda=0.7$, then $m_L$ at 100\hkpc\ rises by
about 0.05 whereas if $\Omega_M=0.2,
\Omega_\Lambda=0.0$, then $m_L=0.50$. The increase
is largely as a result of the rise in the implied co-moving sample
density over this redshift range. 

The merger probability of a kinematic pair depends sensitively on the
pairwise velocity dispersion, $\sigma_{12}$, of galaxies. The model
pairwise velocity distribution is computed as the convolution of the
correlation function with the distribution of random velocities. The
infall velocities are negligible at these small separations and we
will assume that the peculiar velocities are drawn from a Gaussian
distribution.  The measured fraction of the CNOC2 pair sample with
velocities smaller than some $\Delta v$, normalized to the value at
1000\kms, is displayed in Figure~2. The 50\hkpc\ wide pairs limited at
$-19.5$ mag are plotted as open squares, the 20\hkpc\ pairs limited at
$-18.5$ and $-19.5$ mag are plotted as octagons and diamonds,
respectively. The upper curve assumes that $\sigma_{12}$ is 200 \kms\
and the lower one 300
\kms, which approximately span the data.

\section{Merger Rate Estimation}

The merger rate is best estimated from very close kinematic pairs,
20\hkpc\ or less, about half of which are physically close and have
significant morphological disturbance \cite{patton_ssrs2}. However the
fraction of galaxies in such close pairs is about one percent, giving
poor statistics.  Since the number of pairs increases smoothly as
$r^{3-\gamma}$, where $\gamma$ is the slope of the small scale
correlation function, we can use pairs at somewhat larger separations
as statistically representative of the close pairs, however we prefer
to stay within the radius of virialized material around a galaxy over
our redshift range, which is no larger than about 100\hkpc. The mass
accretion rate from major mergers is therefore estimated as,
\begin{equation}
{\cal R}_M = {1\over 2} f_L(\Delta v, r_p,z) C_{zs}(\Delta v,\gamma) 
	F(v<v_{mg})
	\langle M \rangle T_{mg}^{-1}(z,r_p),
\label{eq:mgf}
\end{equation}
where the factor of one half allows for the double counting of pairs,
$C_{zs}(\Delta v,\gamma)$ converts from redshift space to real space pairs, F
gives the fraction of the pairs that will merge in the next $T_{mg}$
(the ``last orbit'' in-spiral time from $r_p$) and $\langle M
\rangle$ is the mean incoming mass as estimated assuming a constant
M/L.  For the relatively massive galaxies considered here the
dynamical friction is so strong that it is more violent relaxation
with little timescale dependence on the masses.  The measured ratio of
the numbers of 50 and 100\hkpc\ pairs to 20\hkpc\ pairs in the CNOC2
sample is $3.8\pm1.0$ and $9.4\pm3.0$, respectively, in accord with
the expectation of a growth as $r_p^{3-\gamma}$ with the inner cutoff
of 5\hkpc.

Not all kinematic pairs are close in physical space.  The relation
between the kinematic pairs closer than $r_p$ and $\Delta v$ and pairs
with a 3D physical distance $r_p$ is readily evaluated by integrating
the velocity convolved correlation function over velocity and
projected radius and ratioing to the 3D integral of the correlation
function.  We find that $C_{zs}=0.54$ for $\Delta v=1000\kms $ and
$\gamma=1.8$ There is support for this value on the basis of
morphological classification, as tested in Patton et al. (1999), where
about half of the kinematic pairs exhibited strong tidal features.

The fraction of physically close pairs that are at sufficiently low
velocity to merge is a key part of the rate calculation.  It is clear
that many galaxies will have close encounters which do not lead to
immediate mergers, although mergers could of course occur on
subsequent orbital passages.  The key quantity that we need is the
ratio of the critical velocity to merge, $v_{mg}$, to $\sigma_{12}$.
The timescale for close pairs to merge is much shorter than the time
over which morphological disturbances are clearly evident, by nearly
an order of magnitude
\citep{bh,mh,dmh}. This is one of our reasons for preferring kinematic
pairs as a merger estimator. The simulation results indicate that the
time to merge is, on the average, roughly that of a ``half-circle''
orbit, which at $r_p$ of 20\hkpc\ at a velocity of 200 \kms\ is close
to 0.3 Gyr. A straight-line orbit with instantaneous merging would
merge in about 0.1 Gyr, although that is not likely to be
representative.

To compute the merger probability, $F(<v_{mg})$, we need to know the
maximal velocity to merge, $v_{mg}$, at a physical separation of
20\hkpc\ for a typical $M_\ast$ galaxy. A not very useful lower bound
is fixed by the Keplerian escape velocity at 20\hkpc, $v_c\sqrt{2}
\kms$, where the circular velocity is approximately 200
\kms. An upper bound to $v_{mg}$ is the velocity that an object would have if it is captured into a galaxy's extended dark halo at the virialization 
radius and orbits to 20\hkpc\ with no dynamical friction. The
virialization radius is approximately at the radius where the mean
interior overdensity is $200\rho_c$, implying $r_{200} = v_c/(10H_0)$,
or, about 200\hkpc\ for our typical galaxy.  The largest possible
apogalactic velocity at $r_{200}$ is $v_c$, which leads to an
undissipated velocity at 20\hkpc\ of $2.37v_c$. Using $\sigma_{12}=
200(300)\kms$ at 20\hkpc, we find that the fraction of all physical
pairs that merge in one $T_{mg}$ is about 0.40(0.16).  Therefore, we
will normalize to a merger probability of 0.3, noting the 50\% or so
uncertainty.

The absolute magnitude limit of $-19.8+5\log{h}$ mag corresponds to
$L\ge 0.5L_\ast$ which contains about 58\% of the luminosity for the
mean CNOC2 LF, $M_\ast=-20.4$ and $\alpha=-1.2$. To make our merger
rate inclusive of major mergers we normalize to $L\ge 0.2L_\ast$,
which includes 85\% of the luminosity. Within the current statistical
accuracy, the paired and field galaxies have identical LFs.  On the
basis of n-body experiments \citep{bh} galaxies with masses greater
than about $0.2M_\ast$ will merge in approximately one orbital time.

On the basis of these considerations we find that the rate of mass
accumulation of galaxies with luminosities of $0.2M_\ast$ and above
is,
\begin{equation}
{\cal R}_M = (0.02\pm0.01)M_\ast (1+z)^{0.1\pm0.5} 
	{F(v_{mg}/\sigma_{12})\over 0.3} {0.3 {\rm Gyr}\over T_{mg} }
	{\rm Gyr}^{-1},
\label{eq:mgr}
\end{equation}
where we have adopted the 100\hkpc\ $m_L$ value for a flat, low
density cosmology and explicitly assumed that the velocity and 
timescale factors do not vary over this redshift range, as expected at
these small scales in a low $\Omega$ universe
\citep{ccc}. There is direct evidence that once evolution compensated
that the luminous galaxies retain have no evolution in their circular
velocities \citep{vogt,gabriella}.

\section{Discussion and Conclusions}

Our main observational result is that for galaxies with $M_R^{k,e}\le
-19.8+5\log{h}$ mag, the fraction of galaxy luminosity in 50\hkpc\
wide kinematic pairs is about 14\%, with no noticeable redshift
dependence over the redshift zero to one range. This implies an
integrated mass accretion rate of about 2\% of $L_\ast$ per Gyr per
galaxy for merging galaxies having $L\ge 0.2L_\ast$.  Our rate is
uncertain at about the factor of two level due to uncertainty in the
dynamical details of merging for our sample definitions.  This merger
rate implies a 15\% mass increase in an $M_\ast$ galaxy since
redshift one. If the correlations of lower luminosity galaxies are
only somewhat weaker than these \citep{roysoc} then the same
$0.15M_\ast$ merged-in mass causes a 50\% mass increase in a
0.3$M_\ast$ galaxy.

There are several issues that require further investigation. First, the
rate of merging of similarly selected kinematic pairs should be
studied in appropriately matched n-body experiments to better
determine the orbital timescales. Second, the absence of a redshift
dependence of $\sigma_{12}$ and $v_{mg}$ needs to
be observationally checked. Third, the connection between close
kinematic pairs and morphologically disturbed galaxies, which does
conform to the kinematic pair predictions at low redshift
\citep{patton_ssrs2}, needs to be better understood at high redshift.

\acknowledgments

This research was supported by NSERC and NRC of Canada.  HL and DWH
acknowledge support provided by NASA through Hubble Fellowship grants
HF-01110.01-98A  and HF-01093.01-97A, respectively, 
awarded by the Space Telescope Science Institute, which is operated by
the Association of Universities for Research in Astronomy, Inc., for
NASA under contract NAS 5-26555.

\newpage
~
\figcaption[fig1.eps]{The fraction of the sample luminosity in pairs
with $\Delta v \le 1000
\kms$ and $5\le r_p\le 50\hkpc$ as a function of redshift. The octagons
are from the CNOC2 sample and the triangles from CFGRS.
\label{fig:flz}}

\figcaption[fig2.eps]{The velocity distribution function for
CNOC2 galaxy pairs. The filled points are for 20\hkpc\ pairs limited
at $M_R^{k,e}$ of $-19.5$ mag (diamonds) and $-18.5$ mag (octagons) and
the open squares are for 50\hkpc\ pairs limited at $-19.5$ mag. The
lines are the distributions expected for an $r^{-1.8}$ correlation
function convolved with Gaussian velocity distribution functions of
200 and 300 \kms, the upper and lower lines respectively.
\label{fig:pv}}

\newpage
\includegraphics[width=0.9\hsize]{fig1.eps} 

\newpage
\includegraphics[width=0.9\hsize]{fig2.eps}

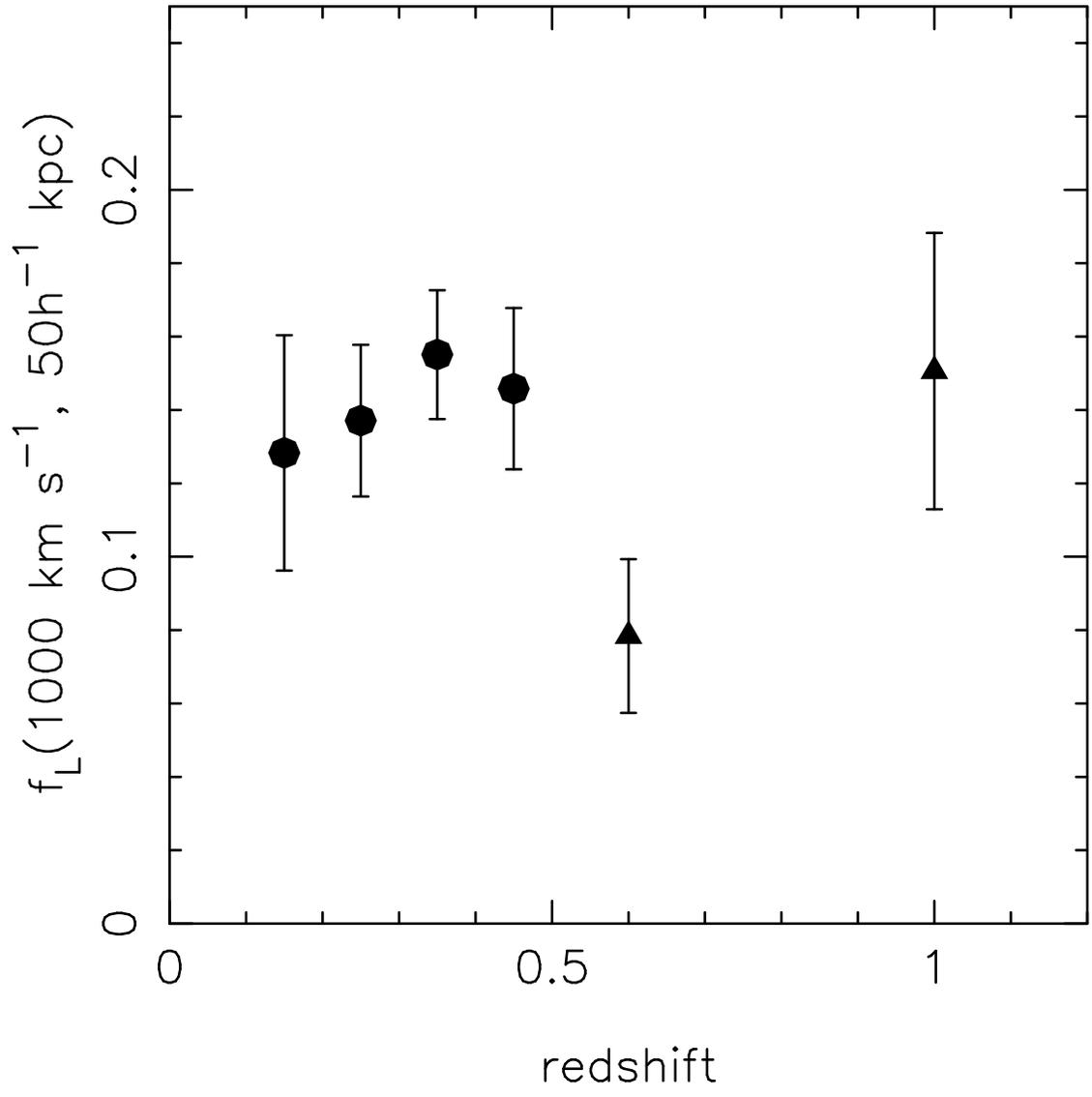
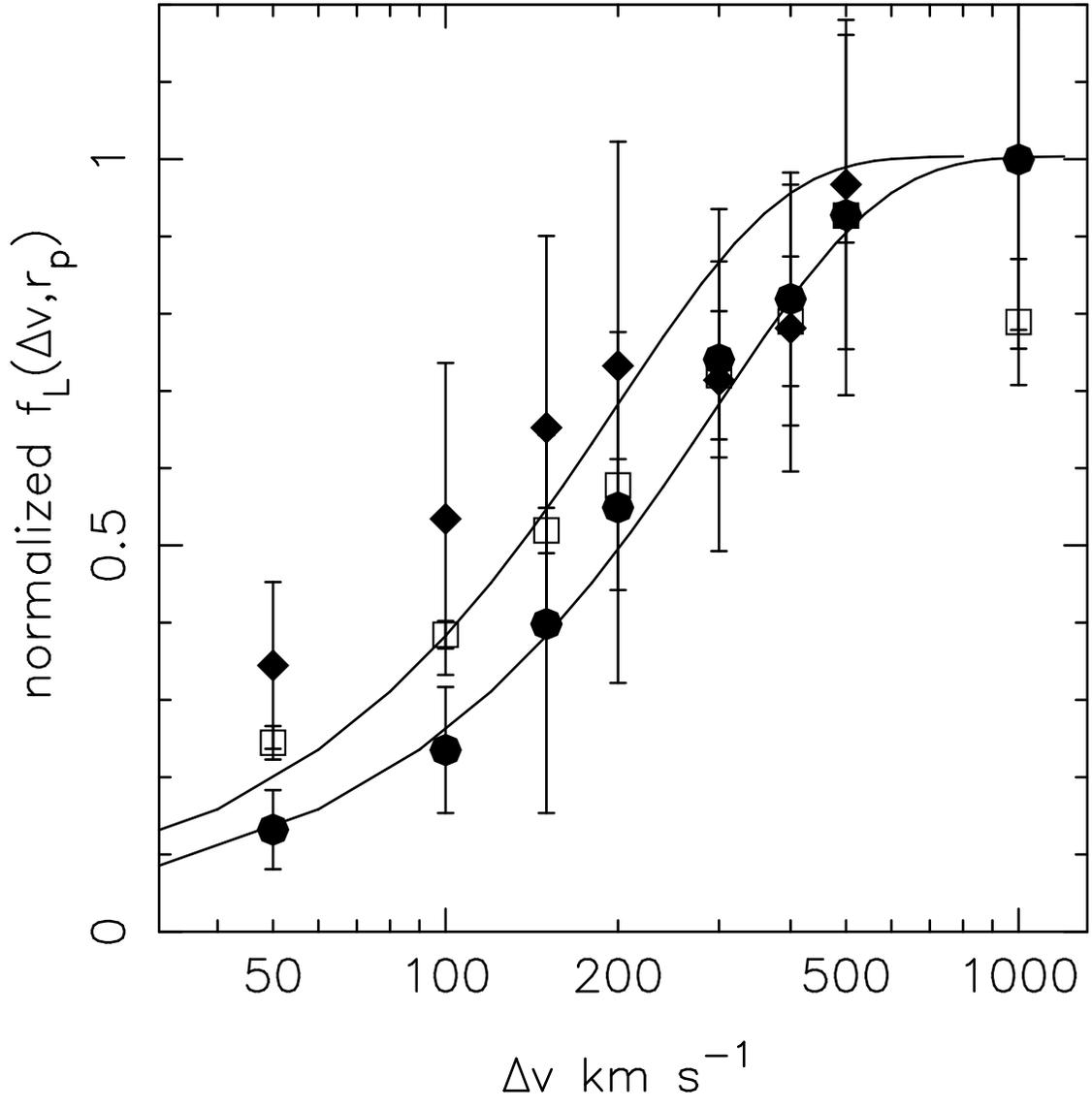
\end{document}